\documentclass[reqno, 11pt]{amsart}              	
\usepackage{hyperref, color, marvosym, enumitem, amsthm}
\definecolor{darkblue}{rgb}{0,0,0.7}
\definecolor{darkgreen}{rgb}{0,0.5,0}
\definecolor{darkred}{rgb}{0.7,0,0}
\hypersetup{breaklinks=true, colorlinks=true, urlcolor=darkblue, linkcolor=darkblue, citecolor=darkblue}
\usepackage{amsmath}
\usepackage{amscd, amsthm, amsfonts, amssymb} 
\usepackage{mathrsfs, bbm, cancel}
\usepackage{capt-of}
\usepackage{graphicx}

\numberwithin{equation}{section}
\newtheorem{proposition}{Proposition}

\theoremstyle{remark}

\theoremstyle{definition}
\newtheorem{definition}[proposition]{Definition}

\newtheorem{remark}[proposition]{Remark}
\newtheorem{example}[proposition]{Example}

\newcommand{\RR}{\mathbbm R}
\newcommand{\NN}{\mathbbm N}

\begin{document}

\title[EMA vs.~MEA]{Exponential moving average versus \\[0.5ex] moving exponential average}
\author{Frank Klinker}
\thanks{{\Large\Letter}: Fakult\"at f\"ur Mathematik, TU Dortmund, 44221 Dortmund, Germany}
\thanks{{\large\Email}\ : \href{mailto:frank.klinker@math.tu-dortmund.de}{frank.klinker@math.tu-dortmund.de}}
\thanks{\vspace*{1ex} Math.~Semesterber. {\bf 58} (2011), no.1, 97-107. \href{http://dx.doi.org/10.1007/s00591-010-0080-8}{DOI 10.1007/s00591-010-0080-8}} 


\begin{abstract} In this note we discuss the mathematical tools to define trend indicators which are used to describe market trends. We explain the relation between averages and moving averages on the one hand and the so called exponential moving average (EMA) on the other hand. We present a lot of examples and give the definition of the most frequently used trend indicator, the MACD, and discuss its properties.\\[2ex]
\end{abstract}
\maketitle


\section{Introduction}\label{intro}

When trends of charts of indices or stocks are investigated then it is common to use so called trend indicators to decide whether to buy, hold or sell. 
But how are these indicators calculated? 
The first idea is, to smooth the curve of points given by the closes and extract informations from the relation between the original and the smoothed curve. 
One assumption on the smoothing, is that local maxima and minima should not be seen by the smoothed curve when they are only short living or of low amplitude. Another assumption is, that the information is achieved from the data alone and not from its development. Therefore, usual simple methods like interpolation by polynomials or Bezier curves are not appropriate, if we do not want to remove single data from our calculations. Another way is to use regression methods. This is not applicable to our problem, too, because the choice of an adequate regression curve needs the knowledge of the shape of the original curve, i.e.\ the development of the data. For the properties of interpolation and regression see for example \cite{num1} and \cite{num2}.
A good way of smoothing turns out to consider averages of the original sequence of closes. But what average should we take? 
In the literature moving averages as well as exponential averages are taken to be good candidates, see for example \cite{appel}, \cite{kauf1}, or \cite{pring}. 

In this note we discuss some frequently used objects such as the so called exponential moving average (EMA), which is not a moving average in a mathematically rigorous sense, and compare it to the rigorous moving exponential average (MEA), see Section \ref{meaea}. Before we do so, we give a precise notion of average and moving average for a real sequence in Sections \ref{av} and \ref{movav} and discuss examples.
In Section \ref{appli} we apply the discussion to trend analysis by introducing the EMA and the moving average convergence/divergence (MACD). We discuss these objects at a basic example, too. 

The motivation for this note came from the request to explain special technical trend indicators like the MACD. This request was annexed a lot of texts from the world wide web. In all of these texts the terms we explain here in a rigorous way were mixed up or were not used properly. Also in the professional literature which explains the handling of trend indicators and discusses markets the authors often do not mention the precise mathematical definitions and structures. Of course, this is caused by the author's intention to describe the market analysis and the system of trading intelligibly to all readers, see e.g.~\cite{PohlBook}.

With this note we give the interested user the chance to learn about the mathematical tools needed to construct trend indicators.

\section{Averages}\label{av}

We consider a sequence of real numbers $x=(x_k)_{k=1,2,\ldots}$ and a further sequence $\alpha =(\alpha_k)_{k=1,2,\ldots}$ where the latter takes its values in the interval $[0,1]$. The aim is to define the average of the sequence $x$ with respect to the weights $\alpha$.
\begin{definition}\label{def1}
Let $x=(x_1,x_2,\ldots )$ be a real sequence and $\alpha=(\alpha_1,\alpha_2,\ldots )$ a sequence taking its values in the interval $[0,1]$.
The $\alpha$-{\em average} of $x$ is defined as the sequence $\delta=(\delta_1,\delta_2,\ldots)$ defined recursively by 
\begin{equation}\label{1}
\delta_{n+1}:=(1-\alpha_{n})\delta_{n}+\alpha_{n} x_{n+1},\ n\geq 1,\qquad \delta_1:=x_1\,. 
\end{equation}
If the limit $\bar\delta:=\lim_{n\to \infty}\delta_n$ exists, then this is called the {\em limit average}. For a finite sequence of length $n_0$ we identify $\bar\delta$ with $\delta_{n_0}$.
\end{definition}

\begin{remark}
This recursively defined element can be expanded to 
\begin{equation}
\delta_n = \prod_{s=\ell}^{n-1}(1-\alpha_s)  \delta_\ell 
			+\sum_{r=\ell+1}^{n}\left(\prod_{s=r}^{n-1}(1-\alpha_s)\right) \alpha_{r-1}x_{r}\,.
\end{equation}
Using this with $\ell=1$ and inserting the initial condition $\delta_1=x_1$ yields
\begin{equation}\label{delta}
\delta_n = \prod_{s=1}^{n-1}(1-\alpha_s)x_1+  \sum_{r=2}^{n}\left(\prod_{s=r}^{n-1}(1-\alpha_s)\right)\alpha_{r-1}x_{r}\,.
\end{equation}
\end{remark}

\begin{definition}
The coefficients $\hat\alpha^{(n)}_s$ in $\delta_n=\sum_{r=1}^n\hat\alpha^{(n)}_r x_r$ are called the {\em weights} of the $\alpha$-average:
\begin{equation}
\hat\alpha^{(n)}_1=\prod_{s=1}^{n-1}(1-\alpha_s),\quad
\hat\alpha^{(n)}_r=\alpha_{r-1}\prod_{s=r}^{n-1}(1-\alpha_s)\,,\quad 2\leq r\leq n\,.
\end{equation}
In particular, $\sum_{s=1}^n\hat\alpha^{(n)}_s=1$.
\end{definition}

\begin{example}\label{ex1} Special choices for $\alpha$ yield well known examples:
\begin{enumerate}[leftmargin=2.25em]
\item[(1)]
The sequence $ \alpha_s=\frac{\mu-\nu+1}{s+\mu+1}$ with $0\leq\nu\leq\mu$ yields
weights
\[\qquad 
\hat\alpha_1^{(n)} =\frac{\binom{n+\nu-1}{n-1}}{\binom{n+\mu}{n-1}},\quad
\hat\alpha_r^{(n)} =\frac{\mu-\nu+1}{r+\mu}\frac{\binom{n+\nu-1}{n-r}}{\binom{n+\mu}{n-r}}\,,\quad 2\leq r\leq n\,.
\]
\begin{itemize}[leftmargin=2.25em]
\item[(a)]
$\mu=\nu=0$ defines the {\em arithmetic mean} with $  \alpha_s=\frac{1}{s+1}$ and  $  \hat\alpha^{(n)}_r=\frac{1}{n}$ as well as $  \delta_n=\frac{1}{n}\sum\limits_{s=1}^n x_s$.
\item[(b)] 
$\mu=1,\nu=0$ defines the {\em weighted arithmetic mean} with $  \alpha_{s}=\frac{2}{s+2}$ and $  \hat\alpha^{(n)}_r=\frac{2r}{n(n+1)}$ as well as  $  \delta_n=\frac{2}{n(n+1)}\sum\limits_{s=1}^n sx_s$.
\end{itemize}
\item[(2)] 
The sequence $\alpha_s=\alpha<1$ yields 
\begin{equation}
\qquad \delta_n=(1-\alpha)^{n-1}x_1 + \alpha\sum_{s=0}^{n-2}(1-\alpha)^s x_{n-s}\,,
\label{EA}\end{equation}
the {\em exponential average (EA) of weight $\alpha$}. Its weights are given by $\hat\alpha^{(n)}_1=(1-\alpha)^{n-1}$ and $\hat\alpha^{(n)}_r=\alpha(1-\alpha)^{n-r}$ for $r\geq 2$. 
\end{enumerate}
\end{example}
\begin{remark}
There are further classical and important averages which are not covered by Definition \ref{def1}. As an example we would like to mention the geometric mean which is defined by $\gamma_n:=\sqrt[n]{x_1x_2\cdot\ldots\cdot x_n}$. The recurrence formula for the geometric mean is given by $\gamma_n=\sqrt[n]{(\gamma_{n-1})^{n-1}\cdot x_n}$. This average is important in applications, for instance the calculation of elliptic integrals uses the geometric mean (see for example \cite{koen}). 
\end{remark}
\begin{remark}\label{rem1}
In the weighted arithmetic  as well as in the exponential average the values of $x$ with higher indices have a bigger influence to the average, due to their higher weights. In particular  the bigger $\alpha$ the less the influence of the lower contributions to the exponential average.

Moreover, because the entries in the weight sequence are constant, the exponential average has the advantage that if we go from $\delta_n$ to $\delta_{n+1}$ we do not need to know how far we have gone in the sequence $x$ to calculate the new average. It is calculated only from the next entry of $x$ and the old average.
\end{remark}

\section{Moving averages}\label{movav}

We modify the discussion from the last section in such a way that we do not consider the whole sequence $x$ up to $n$ to evaluate the average  $\delta_n$. 
As we saw in Example \ref{ex1} (1b) and (2) the  ``older'' the value of $x$ the  smaller the weight with which it enters into the average. 
Therefore we forget about the old ones and for fixed $N$ we let only enter the ``youngest'' ones $(x_{n-N},x_{n-N+1},\ldots, x_n)$ to our new average. This property is sometimes called limited memory. 
\begin{definition}
Let $x=(x_1,x_2,\ldots )$  be an infinite  real sequence and $N\in\NN$ a fixed number. Furthermore let $\alpha=(\alpha_1,\ldots,\alpha_N)$ be a finite sequence taking its values in the interval $[0,1]$. 
For every finite subsequence $x^{(n)}=(x_{n-N+1},\ldots,x_{n})$ of $x$ of length $N$ we define  the associated $\alpha$-average $\delta^{(n),N}\!\!=(\delta^{(n),N}_1,\ldots,\delta^{(n),N}_N) $ as in  Definition \ref{def1}:
\begin{equation}\label{2}
\begin{aligned}
\delta^{(n),N}_{k}&:=(1-\alpha_{k-1})\delta^{(n),N}_{k-1}+\alpha_{k-1} x_{n-N+k},\ \  2\leq k\leq N,\\
\delta^{(n),N}_1&:=x_{n-N+1}\,.
\end{aligned}
\end{equation} 
The $N$-{\em moving} $\alpha$-{\em average} of $x$ is defined by the sequence $\Delta^N=(\Delta^N_1,\Delta^N_2,\ldots)$ with
\begin{equation}
\Delta^N_n:=\delta^{(n),N}_N\,.
\end{equation}
If it exists, we denote the limit by \[\bar\Delta^N:=\lim_{n\to\infty}\Delta_n^N\] and call it the {\em limit moving average}.
\end{definition}

\begin{example} We consider the special choices similar to Example \ref{ex1}.
\begin{enumerate}[leftmargin=2.25em]
\item[(1)] In  case of the arithmetic average, i.e.\ $ \alpha_s=\frac{1}{s+1}$, we get 
\begin{equation*}
\qquad \delta^{(n),N}_k = \frac{1}{k}\big(x_{n-N+1}+\ldots+x_{n-N+k}\big)\,,
\end{equation*}
and the $N$-moving  arithmetic average  is given by
\begin{equation}
\qquad \Delta^N_n = \frac{1}{N}\big(x_{n-N+1}+\ldots + x_{n}\big)\,.
\end{equation}
\item[(2)]  In case of the weighted arithmetic mean, i.e.\  $ \alpha_s=\frac{2}{s+2}$,  we get 
\begin{equation*}
\qquad \delta^{(n),N}_k = \frac{2}{k(k+1)}\big( x_{n-N+1}+2x_{n-N+2}+\ldots+k x_{n-N+k}\big)\,,
\end{equation*}
and the $N$-moving arithmetic mean is given by
\begin{equation}
\qquad \Delta^N_n = \frac{2}{N(N+1)}\big(x_{n-N+1}+2x_{n-N+2}+\ldots + N x_{n}\big)\,.
\end{equation}
\item[(3)] In case of the exponential average of weight $\alpha$ we get
\begin{equation*}
\qquad \delta^{(n),N}_k = (1-\alpha)^{k-1}x_{n-N+1}+\alpha\sum_{s=0}^{k-2}(1-\alpha)^s x_{n-N+k-s}\,,
\end{equation*}
and the $N$-moving exponential average (MEA) is given by
\begin{equation}
\qquad \Delta^N_n = (1-\alpha)^{N-1}x_{n-N+1}+\alpha\sum_{s=0}^{N-2}(1-\alpha)^s x_{n-s}\label{MEA}\,.
\end{equation}
\end{enumerate}
\end{example}

\begin{example}
\begin{enumerate}[leftmargin=2.25em]
\item[(1)]
In our first example we consider the sequence $x$ with $x_s=s$. Then  $\delta^{(n),N}_k$ as well as the $N$-moving averages $\Delta^N_n$  are given in terms of $\delta_n$ by
\[
\qquad \delta^{(n),N}_k=(n-N)+ \delta_k , \quad  \Delta^N_n =(n-N)+\delta_N
\]
with
\begin{itemize}[leftmargin=2em]
\item arithmetic mean
\[
\displaystyle \delta_n = \frac{n+1}{2}
\]
\item weighted arithmetic mean
\[
\displaystyle \delta_n = \frac{2n+1}{3}
\]
\item exponential average
\[
\displaystyle \delta_n = n-\frac{1-\alpha}{\alpha}\big(1-(1-\alpha)^{n-1}\big)
\]
\end{itemize} 
\item[(2)]
The second example deals with the sequence defined by $ x_s=\frac{1}{s}(1-\beta)^s$. This yields the following averages\footnote{This example has been corrected compared with the published version.}:
\begin{itemize}[leftmargin=2em]
\item
arithmetic mean
\[
\displaystyle\delta_n \approx \frac{1}{n}\Big(\ln\frac{1}{\beta}-\frac{1}{n+1}\Big)
\]
\item
weighted arithmetic mean
\[
\displaystyle \delta_n = \frac{ 2 (1-\beta) (1-\beta^{n+1}) }{ n(n+1)\beta }
\]
\item
exponential average
\[
\displaystyle \delta_n \approx (1-\alpha)^n\Big(1-\beta-\frac{\alpha}{n+1}+\alpha\ln\frac{1-\alpha}{\beta-\alpha}\Big)
\]
\end{itemize}
Here we used the expansion \[\qquad  \ln\frac{1}{1-x}=\sum\limits_{k=1}^\infty\frac{x^k}{k}\approx \sum\limits_{k=1}^n\frac{x^k}{k}+\frac{1}{n+1}\,.\]
\end{enumerate}
\end{example}

\section{Comparing MEA and EA}\label{meaea}

To compare the exponential average \eqref{EA} and the moving exponential average \eqref{MEA} and their respective limits, we turn from the finite sequence $x^{(n)}=(x_1,x_2,\ldots,x_n)$ to the finite sequence $y^{(n)}=(y^{(n)}_0,\ldots,y^{(n)}_{n-1})=(x_{n},x_{n-1},\ldots,x_1)$, i.e.\ $y^{(n)}_k:=x_{n-k}$. This numbering ensures, that the ``youngest'' element of the sequence belongs to the lowest index. Then \eqref{EA} and \eqref{MEA} are rewritten as 
\begin{equation*}
\delta_n=(1-\alpha)^{n-1}y^{(n)}_{n-1} +\alpha\sum_{s=0}^{n-2}(1-\alpha)^sy^{(n)}_{s} \tag{\ref{EA}'}
\end{equation*}
and 
\begin{equation*}
\Delta^N_n = (1-\alpha)^{N-1}y^{(n)}_{N-1}+\alpha\sum_{s=0}^{N-2}(1-\alpha)^s y^{(n)}_{s} \tag{\ref{MEA}'}
\end{equation*}
for any finite subsequence and look quite similar. 

Now let $x=(x_1,x_2,x_3,\ldots)$ be an infinite sequence with finite subsequences $x^{(n)}$ and associated  inversed sequences $y^{(n)}$. Then the limit exponential average as well as the limit moving exponential average of the sequence $x$ is\footnote{These expressions seem to be a little formal, because we deal with terms like $y_s:=x_{\infty-s}$, i.e.\ we need to know the whole starting sequence in reverse order. In practical applications this is not a problem, because the sequences are given in the way ``the first data is the youngest one''.}
\begin{equation}\begin{aligned}\label{formally}
\bar\delta &=\alpha \sum_{s=0}^{\infty}(1-\alpha)^s y_{s}\,,\\
\bar\Delta^{N} &= (1-\alpha)^{N-1}y_{N-1}+\alpha\sum_{s=0}^{N-2}(1-\alpha)^s y_{s}\,.
\end{aligned}\end{equation}
To compare EA and MEA, we calculate the difference
\begin{equation}\label{4}
\begin{aligned}
\bar\delta-\bar\Delta^{N} 
	&= \alpha\sum_{s=N-1}^\infty(1-\alpha)^sy_{s}-(1-\alpha)^{N-1}y_{N-1}\\
&=(1-\alpha)^{N}\left( \alpha\sum_{s=0}^\infty(1-\alpha)^sy_{s+N} -y_{N-1}\right)
\end{aligned}
\end{equation}
\begin{definition}\label{adm}
A sequence $x$ is called admissible, if it does not spread widely. More precisely, for $N\in\NN$ we assume $\alpha\displaystyle \sum_{s=0}^\infty(1-\alpha)^sy_{N+s}\approx \bar\delta$ and $\displaystyle\frac{|\bar\delta-y_{N}|}{\bar\delta}<1$.
\end{definition}
From the calculations in (\ref{4}) we get the following observation.
\begin{proposition}
If the initial sequence is admissible then 
\begin{equation}\label{app}
\frac{|\bar\delta-\bar\Delta^{N}|}{\bar\delta} < (1-\alpha)^{N}\,.
\end{equation}
\end{proposition}
In particular, formula (\ref{app}) tells us, that for a given sequence -- under the assumptions of admissibility -- the relative error by replacing the limit EA by the limit MEA only depends on the moving length $N$. 

\begin{remark}
Sometimes the value which we get by canceling the sum of $\bar \delta$ in (\ref{formally}) is taken as an average and the error is estimated with regard to this finite sum. The problem is, that this value is not an average in the sense of Definition \ref{def1}, because the weights do not sum up to 1. Nevertheless, under the assumption of admissibility, the error will be the same.
\end{remark}

The inversion of the sequence $x$ is also needed, when we want to add new values to a given sequence and compare their respective limit averages. In this situation, which will be important in the application in the next section, we start with an infinite (inversed) sequence $y^{[k]}=(y^{[k]}_0,y^{[k]}_1,\ldots)$ at a ``day'' $k$. The next day $k+1$ we add an element to get a new sequence expanding the old one. The new element should be the initial value of our new sequence, and so we call it $y^{[k+1]}_0$. The whole sequence $y^{[k+1]}=(y^{[k+1]}_0,y^{[k+1]}_1,\ldots)$ is then supplemented by $y^{[k+1]}_i:=y_{i-1}^{[k]}$ for $i\geq 1$. 

For finite sequences $y^{(n)}$ and $y^{[n]}$ coincide, when we define $y^{[n+1]}_0:=x_{n+1}$ and $y^{[1]}:=(x_1)$. The inversed sequence $y$ of the infinite sequence $x$ can then be viewed as the limit sequence of $y^{[k]}$.

\section{Application to index charts}\label{appli}

When we want to discuss the chart of a stock or an index it is necessary to smooth the curve, given by the closes. This may be done by taking an appropriate average instead of the closes itself. One condition which can be taken as natural is that the ``younger'' closes are more important than the ``older'' ones. Therefore, the weighted arithmetic and the exponential average seem to be proper candidates. One obvious advantage of the exponential average is that the sequence $\alpha$ is constant (see Remark \ref{rem1}). 
More restrictively we could demand that the market do not see very old closes at all but only closes of a specific amount of periods. 
This leads us to the moving variants of the averages, where we always consider a fixed amount of closes. 
But in the context of moving averages we lose the nice recurrence formula (\ref{1}), because the initial condition changes every day, see (\ref{2}). Nevertheless, these averages have their application in trend analysis, too, see \cite{pring}.

As we saw in (\ref{app}) the relative error we make when we pass from the moving exponential average to its non moving variant is given by $(1-\alpha)^{N}$. This is true, when we make some mild assumptions on the starting sequence $x$ (see Definition \ref{adm}). 
A usual way to model the moving length is the choice $\alpha=\frac{\rho}{N+1}$ with suitable $\rho\in\RR$. For instance $\rho=2$ yields a relative error less than $13.5\%$, and an error less than $1\%$ is achieved by $\rho\geq 4.7$. This can be seen by using $(1-\frac{\rho}{N+1})^N\leq e^{-\rho}$.  The moving exponential average with this special choice of weight is called $N$-day EMA in the literature, see e.g.\ \cite{alex}, \cite{appel}. 

\begin{definition}\label{close} 
The values of an index (or an stock) at the closes of the market define a sequence of non-negative real numbers $c^{[k]}=(c^{[k]}_0,c^{[k]}_1,\ldots )$ and is called the sequence of closes at day $k$. The value $c^{[k]}_0$ is called the close at day $k$. The sequence of closes obey 
$c^{[k]}_i=c^{[k-1]}_{i-1}$
 for $i\geq 1$, such that $c^{[k]}_0$ is always the youngest contribution.
\end{definition}

\begin{definition}\label{def3}
The $N${\em -day exponential moving average (EMA)}  at close $c^{[k]}$ is defined by the exponential average of weight $\frac{2}{N+1}$ of the series $c^{[k]}$ and will be denoted by $E_k^N(c)$, i.e.
\begin{equation}
E_k^N(c) := \bar\delta^{\text{EA}}\Big|_{\alpha=\tfrac{2}{N+1}, y=c^{[k]}}\,.\label{EMA}
\end{equation}
\end{definition}
Within the investigation of market trends we use a further EMA, which is based on a sequence defined by two EMAs of the sequence of closes. The so constructed EMA is used to produce a trend indicator for the initial sequence of closes.
\begin{definition}\label{def4}
The {\em MACD (Moving Average Convergence/Divergence)} at the close $c^{[k]}$  associated to the moving lengths $N_1$ and  $N_2$ is defined as the difference 
\begin{equation}
M^{N_1,N_2}_k(c):=E^{N_1}_k(c)-E^{N_2}_k(c)\,. 
\end{equation}
The MACD defines a new sequence $m^{[k]}$ with $m^{[k]}_0:=M^{N_1,N_2}_k$ and $m^{[k]}_i:=m^{[k-1]}_{i-1}$. 
We call the close $c^{[k]}$ (or the day $k$) $N_0${\em -short} or $N_0${\em -long} if the difference $m^{[k]}_0-E_k^{N_0}(m)$ is negative or positive, respectively.
\end{definition}
\begin{remark}
The change from short to long and from long to short may be taken as indicator for ``buy'' and ``sell'', respectively.
\end{remark}
\begin{example}\label{ex}
Figures \ref{f1} and \ref{f2} show an application of Definition \ref{def3} to the values of the MDAX between December 8, 2008 and April 3, 2009.\footnote{The closes from Example \ref{ex} were kindly provided by \href{http://www.easytrend24.de}{\bf EasyTrend24}.}
We used  a short term EMA ($N_1=12$) and a minor intermediate EMA ($N_2=26$) to define a very short term EMA of the MACD ($N_0=9$). This is a common combination used in market analysis, see \cite{kauf1}. This combination yields two sell and two buy signals.
\end{example}

\section{Concluding remarks} 

The EMA, as used in Section \ref{appli} as well as in the literature, is not a moving average in the mathematical sense of Section \ref{movav}. In fact, it is a usual limit exponential average with a specific weight. For its calculation we do not use a finite subsequence of the closes $c$ of constant length, but all of it. On the one hand, the name is motivated my the special choice of the weight and the error discussion above, see (\ref{app}). On the other hand a further motivation for the term moving comes from the fact that $c^{[k]}$ is by construction an infinite subsequence of $c^{[k+1]}$, see Definition \ref{close}.

As noticed in Example \ref{ex} and the subsequent remark the MACD and its EMA are used to define ``buy'' and ``sell'' signals.
For small $N$  local effects in $c$ play a major role, so that the incidents buy/sell will appear more often. In particular if the sequence of closes itself is range-bound, then maxima and minima within this region lead to unnecessary signals (compare Example \ref{ex} with figures \ref{f3} and \ref{f4} in particular regions day 1 to day 13 and day 64 to day 75). It is a question of  strategy which moving lengths we take in Definition \ref{def3} and \ref{def4}, or if we -- in particular for small $N$ -- replace the EMA by another moving average. 

Another way to handle the sometimes unnecessary signals in range-bound regions is the introduction of more subtle indicators such as the ADX (Average Directional Movement Index) or the RSI (Relative Strength Index). For the calculation of the ADX -- in contrast to the RSI -- the data sequence we used in Section \ref{appli} are not sufficient. In addition we need the highest and lowest value within one period. For more details on this see for example \cite{adx-lit}.

The MACD is a trend follower, i.e.\ the signals always appear after the extrema -- how far depends on the choice of the parameters. In this sense the trend followers only describe the direction of the trend of the data but neither describes the strength of the trend nor does it make predictions on the future development of the data. The first problem may be weakened by introducing indicators like the ADX, as noticed before. One method to gain more information about the future development is the introduction of a trend forecasters, see for example \cite{Mendel}. Usually this is made by using regression curves to extrapolate the closes, e.g.\ linear, polynomial, logarithmic or exponential regression, see \cite{kauf2} or \cite{num2}. Which method is used depends highly on the data structure. Therefore the calculation of such trend forcasters -- in contrast to the trend follower -- needs more than the pure data but a good knowledge of its developing.

\begin{center}
\begin{minipage}[t]{\linewidth}
\centering
\includegraphics[width=\textwidth]{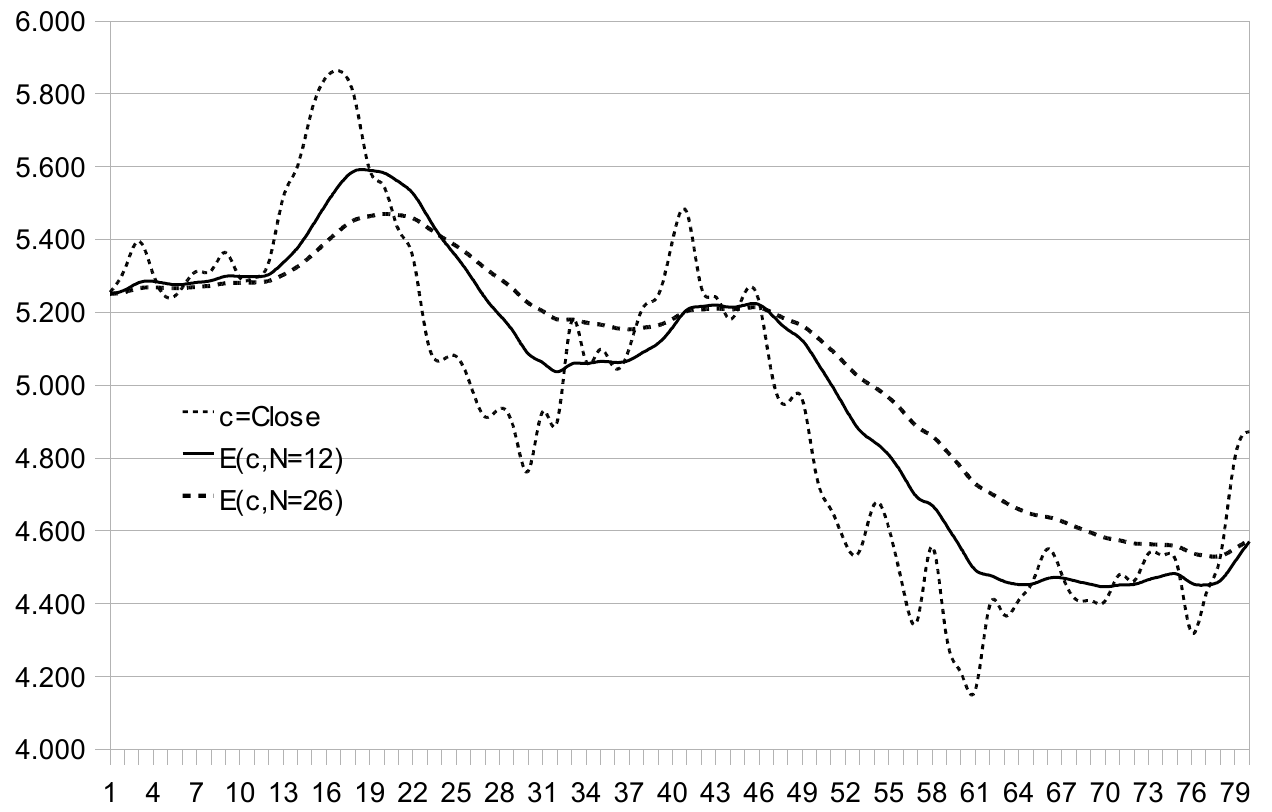}\\[2ex]
\captionof{figure}{\small Series of closes, $c$, and its 12-day and 26-day EMA, $E^{12}(c)$ and $E^{26}(c)$, for the period 2008-12-08 to 2009-04-03.}\label{f1}

\centering
\includegraphics[width=\textwidth]{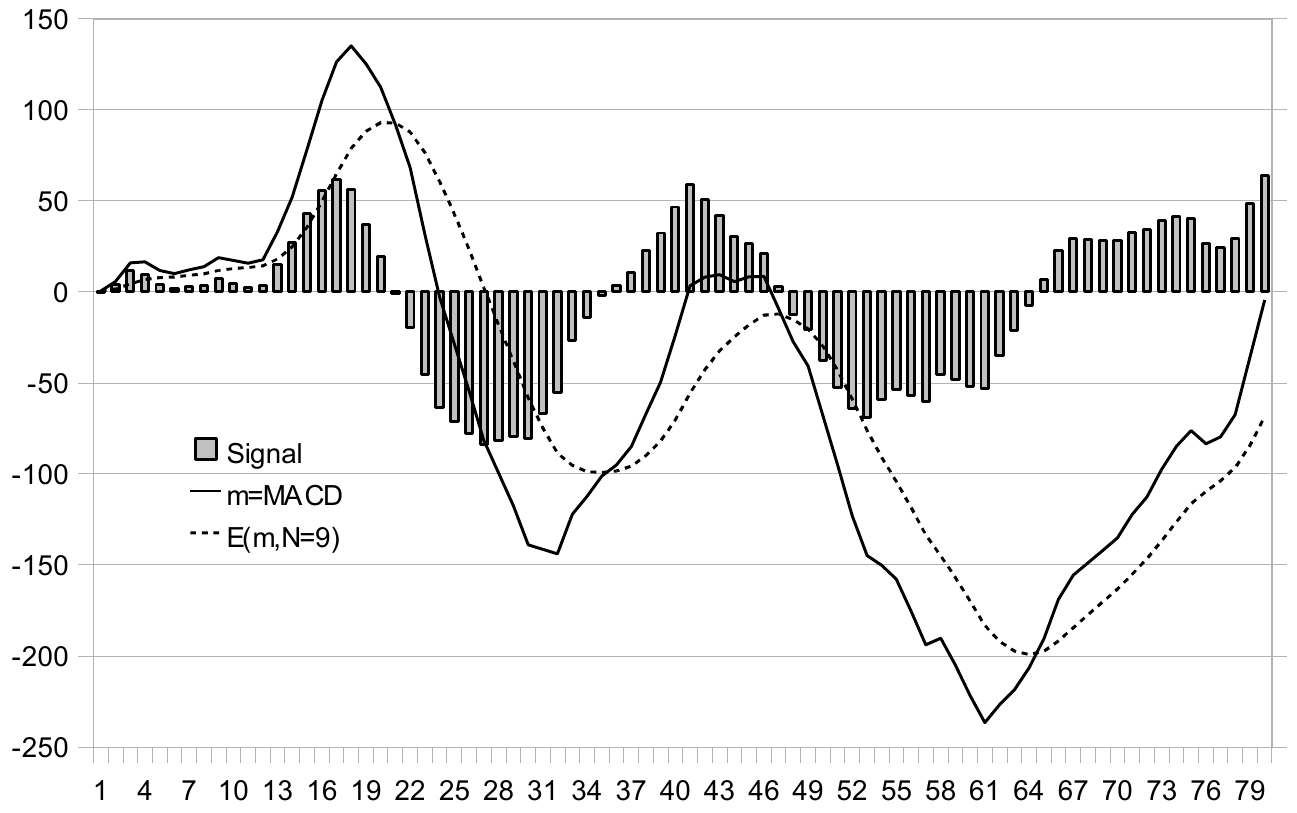}\\[2ex] 
\captionof{figure}{\small MACD, $m=M^{12,26}(c)$, its 9-day EMA, $E^9(m)$, and the associated signal for the period 2008-12-08 to 2009-04-03.} \label{f2}
\end{minipage}
\end{center}

\begin{center}
\begin{minipage}[t]{\linewidth}
\centering
\includegraphics[width=\textwidth]{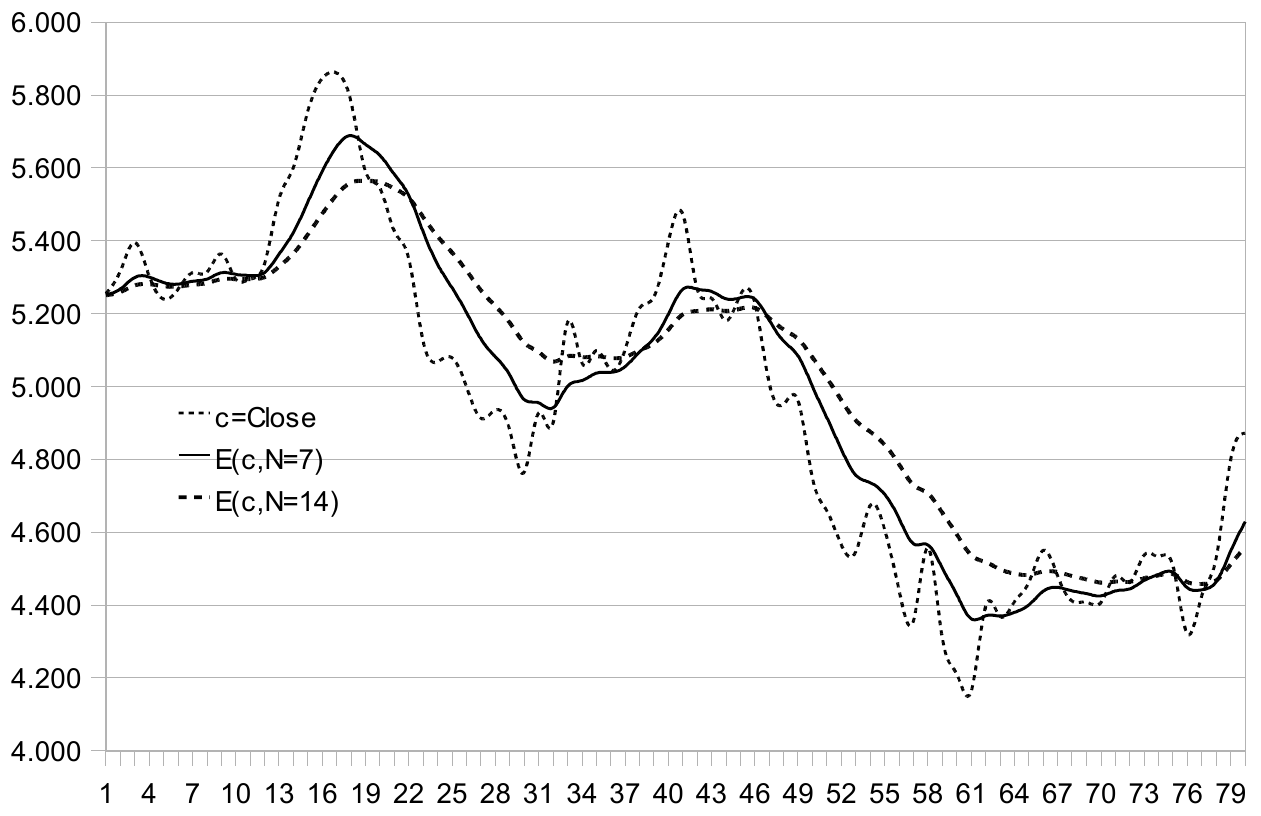}\\[2ex]
\captionof{figure}{\small Series of closes, $c$, and its 7-day and 14-day EMA,\ \ $E^7(c)$ and $E^{14}(c)$, for the period 2008-12-08 to 2009-04-03.}\label{f3}

\centering
\includegraphics[width=\textwidth]{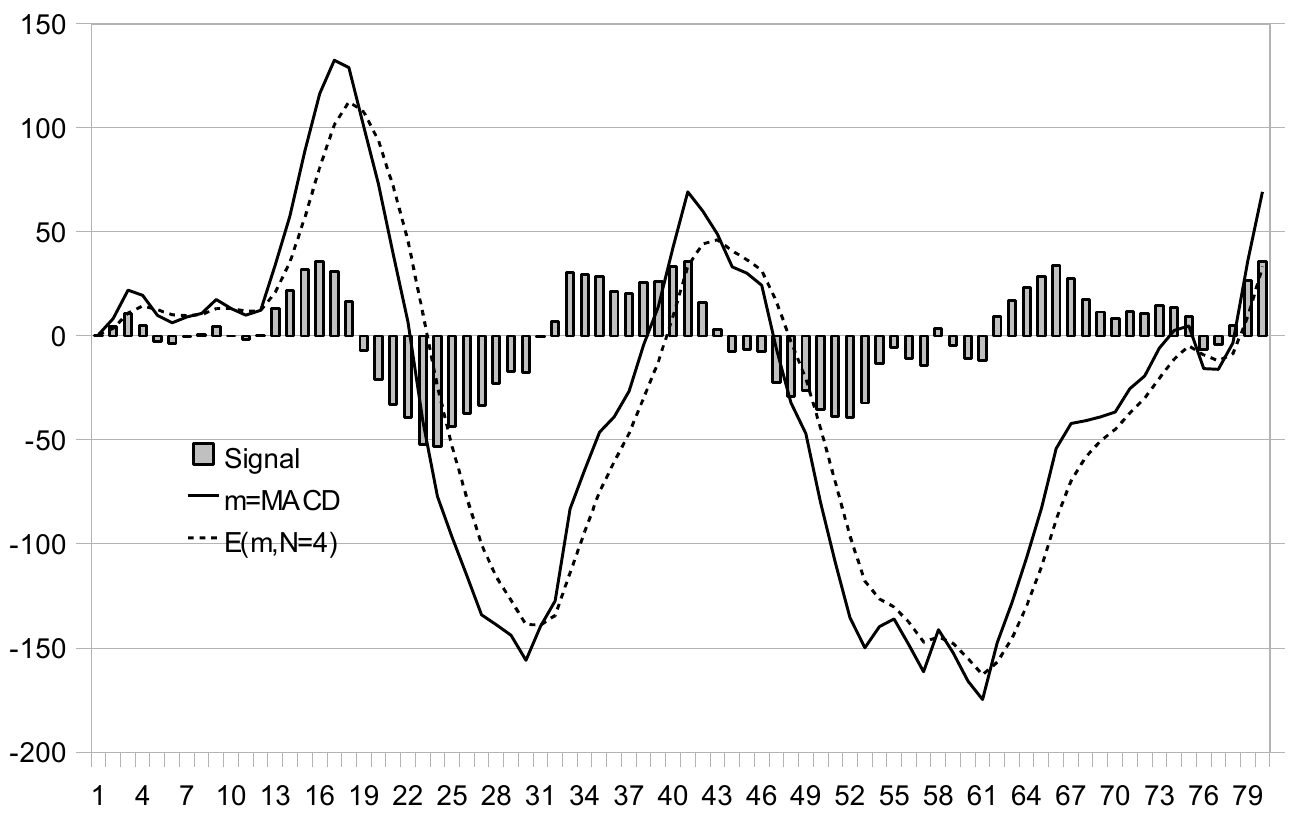}\\[2ex]
\captionof{figure}{\small MACD, $m=M^{7,14}(c)$, its 4-day EMA, $E^4(m)$, and the associated signal for the period 2008-12-08 to 2009-04-03.}\label{f4}
\end{minipage}
\end{center}

\end{document}